\documentclass[final]{ias2}

\usepackage{graphicx} 
\usepackage{multirow}
\usepackage{array}

\usepackage{caption}
\usepackage{subcaption}

\usepackage{hyperref} 

\begin{document}

\markboth{growth of dendritic drying patterns}{K. M. Kolwankar et al.}

\title{Effect of heat source on the growth of
dendritic drying patterns}

\author[rjc]{Kiran M. Kolwankar}
\author[mudp]{Pulkit Prakash} 
\author[mudp]{Shruthi Radhakrishnan} 
\author[mudp]{Swadhini Sahu}
\author[tifr]{Aditya K. Dharmadhikari}
\author[mu]{Jayashree A. Dharmadhikari} 
\author[tifr]{Deepak Mathur} 
\address[rjc]{Department of Physics, Ramniranjan Jhunjhunwala College, Ghatkopar(W), Mumbai 400 086, India}
\address[mudp]{Department of Physics, University of Mumbai, Vidyanagari, Santacruz East, Mumbai 400 098, India}
\address[tifr]{Tata Institute of Fundamental Research, 1 Homi Bhabha Road, Mumbai 400 005, India}
\address[mu]{Department of Atomic and Molecular Physics, Manipal University, Manipal 576104, India}

\begin{abstract}

Shining a tightly-focused but low-powered laser beam on an absorber dispersed in a
biological fluid gives rise to spectacular growth of dendritic patterns. These result from localized drying of the fluid because of efficient absorption and conduction of optical energy by the absorber. 
We have carried out experiments in several biologically relevant
fluids and have analyzed patterns generated by different types of absorbers.
We observe that the growth velocity
of branches in the dendritic patterns can decrease below the value expected for natural drying.
\end{abstract}

\keywords{Dendritic solidification, branched structures, diffusion limited growth}

\pacs{47.53.+n, 47.20.Hw, 47.54.De}
 
\maketitle


\section{Introduction}

Various kinds of patterns are found in nature and they have attracted interest
for ages. A subset of such patterns consists of branched dendritic patterns on which we report in the following. Though dendritic patterns have been observed and experimentally well studied, their theoretical understanding has begun to emerge only relatively recently~\cite{BJH,Lan,KKL,BM,Son}.
Even now there is no complete understanding of various patterns arising in systems that are out-of-equilibrium, specifically the formation of branched structures, viscous fingering and dendritic solidification. In recent years dendritic growth has also been observed in organic molecules~\cite{KHV} and in Langmuir monolayers~\cite{FCGC}.

In dentritic solidification of a pure 
substance from its supercooled melt the formation of patterns is driven by
supercooling and is controlled by the diffusion of the latent heat that is generated
in the transformation away from the interface. This usually leads to dendritic
pattern with branching instability. In the well-known
Hele-Shaw cell  two immiscible fluids are constrained to move between
narrowly separated parallel plates. The viscous fingering 
patterns arise in this cell when high viscosity fluid
is displaced by low viscosity fluid under pressure.
Under normal circumstances the patterns show tip splitting instability.

We have conducted experiments in which we observe dendritic patterns in various biological fluids when low power laser light irradiates an absorber suspended in such medium. For appropriate lair wavelengths, the absorber absorbs the incident optical energy and, as a result, heat is transferred to the liquid medium. In other words, the laser irradiated absorber acts as a heat source. If the absorber is small in size (tens of $\mu$m diameter), and the laser beam is tightly focused (also to tens of $\mu$m diameter), a highly localised region of dehydration can be formed around the absorber, giving rise to an interface between a region that is essentially dry and the liquid medium. This interface
becomes unstable and results in a branched structure, a manifestation of the famous Mullins-Sekerka instability~\cite{mullins}. The formation of the 
patterns that result seem to be reversible in the sense that once the laser is switched
off, the structure dissolves in the liquid - unless the liquid itself is on the verge of drying.

In a recent study~\cite{Bas} we reported on the formation of such patterns in BSA (bovine serum albumin) suspended in PBS (phosphate buffer solution). We discovered accelerated growth of dendrites even for very small laser powers (a few mW). We measured the tip radius and tip velocity and found that though the product $vr^2$
eventually becomes constant, as predicted by the microscopic solvability theory,  it deviates significantly from the constant value on very short time scales. We also observed that the velocity decreased with time, as expected, but, unexpectedly, we found that the velocity can decrease below the value that is expected for natural drying (that is, when patterns are formed in the absence of a laser beam, by normal evaporation). This counterintuitive observation is the central focus of our present study in which we report results of a more systematic investigation that appears to confirm our earlier observations.

\section{Experimental details}

We have carried out experiments in (i) PBS+Agarose (ii)PBS+BSA, and (iii) PBS+Gelatinein solutions. The different solutions used in our experiments were prepared in the following manner: 1-2\% (w/v) Agarose, BSA and gelatin was dissolved in 300 mOsm PBS.A small quantity (0.5-1 mg) of an absorber (of 1064 nm light) was added to each solution and then dispersed by ultrasonication (for 15 s and 5 cycles). A small quantity of prepared solution was placed in a thin (0.1 mm) glass cell and was irradiated by a continuous wave (cw), 1064 nm wavelength Nd:YVO$_4$ laser beam that was tightly focused (to $\sim$1 $\mu$m diameter spot) with a 100X microscope objective (numerical aperture 1.3). Details of our experimental apparatus have been published in recent reports on bubble dynamics \cite{earlier1,earlier2}. 

The patterns that we observed upon irradiating each sample were imaged onto a CCD camera (JVC TKC-1480E operating at 25 Hz) coupled to a computer that recorded the observations in real time. 

We used single-walled carbon nanotubes (SWNT), multi-walled carbon nanotubes (MWNT), and graphite as absorbers. The absorption coefficients for 1064 nm light are all very large: SWNT (10$^5$ cm$^{-1}$), MWNT (8$\times$10$^3$ cm$^{-1}$), graphite (5$\times$10$^5$ cm$^{-1}$) \cite{jpc}. The thermal conductivity of SWNT (3500 W mK$^{-1}$) is far in excess of values possessed by MWNT (50-150 W mK$^{-1}$) and graphite (80-240 W mK$^{-1}$) \cite{jpc}. On the basis of these numbers it would be expected that SWNTs would be the best of the three mediators; our experiments, indeed, confirmed that SWNT-mediated growth was most efficient in terms of accelerated growth.

\section{Theoretical details}

To place our observations of viscous fingering and dendritic solidification in proper context we note that the Laplace equation for pressure in a fluid describes the problem of 
viscous fingering while the diffusion equation for the temperature field 
describes the dendritic solidification (see, for instance,~\cite{KKL} for a detailed review).
In our experiments the liquid-solid boundary itself is moves and, consequently, choice of the proper boundary condition is of utmost importance because it is the boundary condition that introduces the nonlinearity in the problem and also decides the velocity with which the interface moves. The Laplace and diffusion equations may also be replaced by equivalent integral equations. Several analytical~\cite{KKL} and numerical methods~\cite{BWBK,PK,Pro} have been developed to study these systems. Earlier attempts used a perturbative approach but it is only relatively recently that the singular nature of surface tension has become clear, necessitating the development of so-called microscopic solvability theory and recognition of the crucial importance of anisotropy.

The equations governing the heat flow in the two phase problem
are given by~\cite{Cra}
\begin{eqnarray}
c_i \rho_i {{\partial u_i}\over{\partial t}}
= K_i \nabla^2u_i + Q_i,
\end{eqnarray}
where $i=1$ refers to the solid phase and $i=2$ refers to
the liquid phase. $u_i$ denotes temperatures in the two phases and
$c_i$, $\rho_i$, $K_i$ denote, respectively, the specific heat, density and heat conductivity. We assume these parameters to be the same in both phases. $Q_i$ is the heat generation term. In our
setting $Q_1$ is taken to be a delta function at the origin and $Q_2=0$. At the interface we have Stefan's boundary condition, which expresses the heat balance on the interface
\begin{eqnarray}\label{eq:bc}
-L\rho v_n = K_2 {{\partial u_2}\over{\partial n}}
- K_1 {{\partial u_1}\over{\partial n}},
\end{eqnarray}
where $n$ is the outward normal to the moving boundary and $v_n$ 
is the velocity of this boundary along the normal. We note that the heat term makes the equation inhomogeneous.

One can replace this system of equations by an equivalent integral
equation~\cite{KKL} which gives the temperature at any point as
\begin{eqnarray}
u({\bf x},t) = u_\infty + &&\int dS' \int^t_{-\infty} dt' G({\bf x}-{\bf x'},t-t')Lv_n({\bf x'},t)/c
\nonumber \\
&+& \int dV' \int^t_{-\infty} dt' G({\bf x}-{\bf x'},t-t')Q({\bf x'},t)/c\rho,
\end{eqnarray}
where $dS'$ and $dV'$ are the surface and volume elements, respectively, and
\begin{eqnarray}
G({\bf x},t) = (4\pi Dt)^{-d/2} \exp (-{\bf x}^2/4Dt).
\end{eqnarray}
Here one can make simplifying assumptions that $Q$ is a delta function
at the origin and is also taken to be independent of time.
Analysis of these equations is expected to offer a theoretical overview of the 
system under study. Some insights that pertain to our experimental observations are discussed
in the next section.

\section{Results}\label{sec:res}

\begin{figure}
\centering
\begin{subfigure}{\textwidth}
\includegraphics[width=\textwidth]{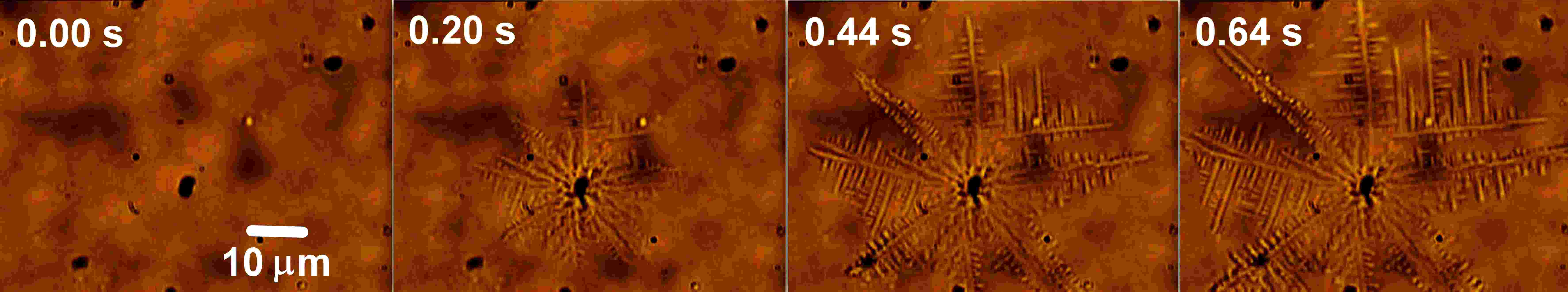}
\caption{}
\end{subfigure}
\begin{subfigure}{\textwidth}
\includegraphics[width=\textwidth]{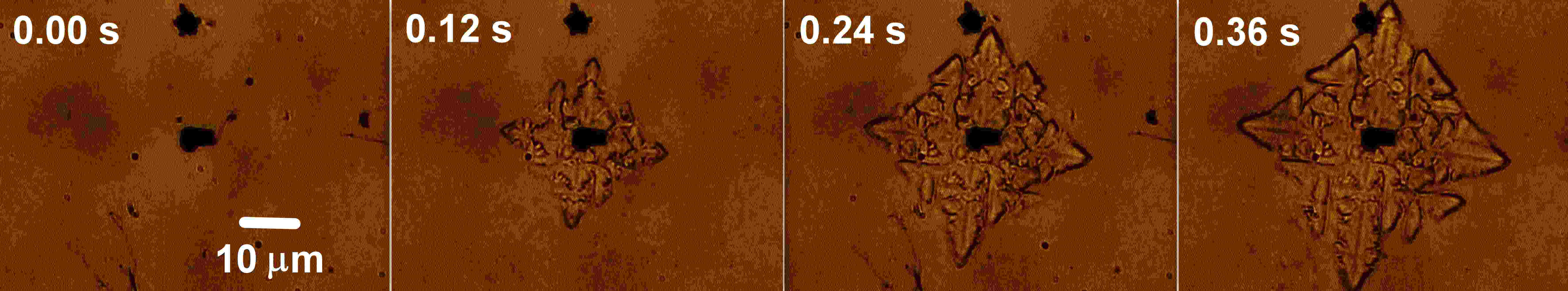}
\caption{}
\end{subfigure}
\begin{subfigure}{\textwidth}
\includegraphics[width=\textwidth]{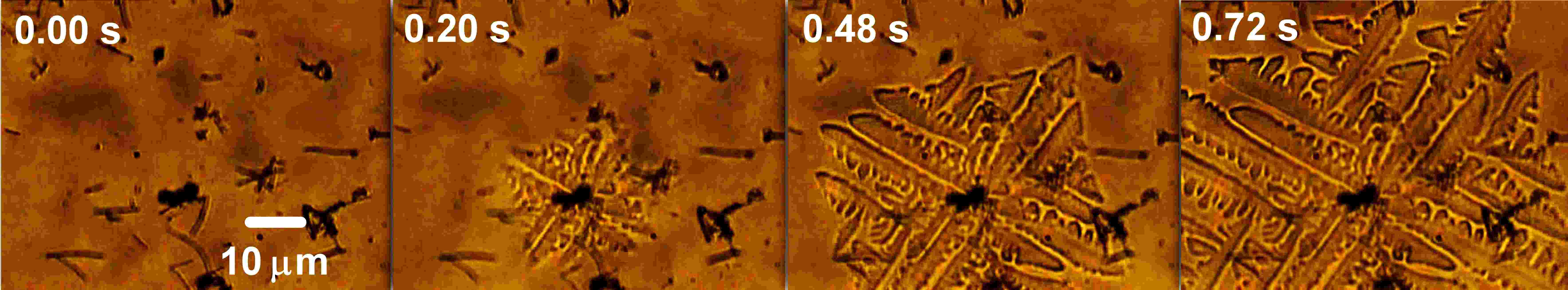}
\caption{}
\end{subfigure}
\begin{subfigure}{\textwidth}
\includegraphics[width=\textwidth]{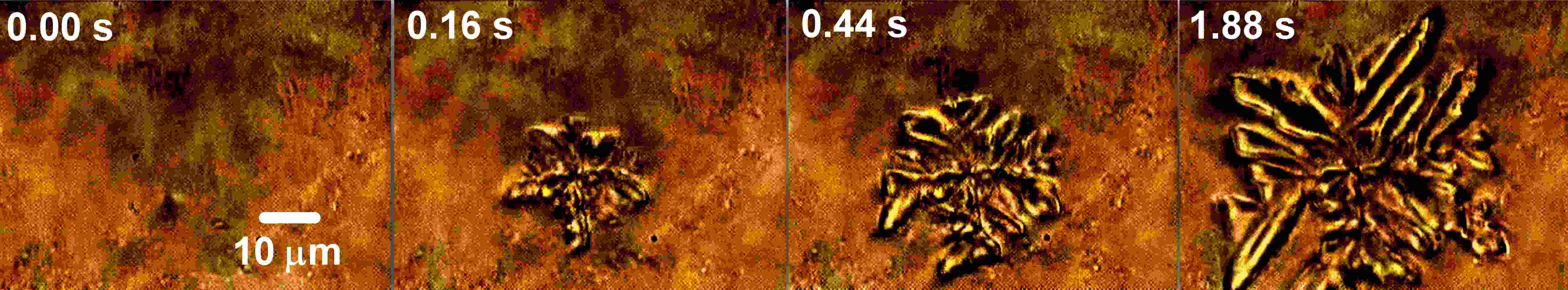}
\caption{}
\end{subfigure}
\begin{subfigure}{\textwidth}
\includegraphics[width=\textwidth]{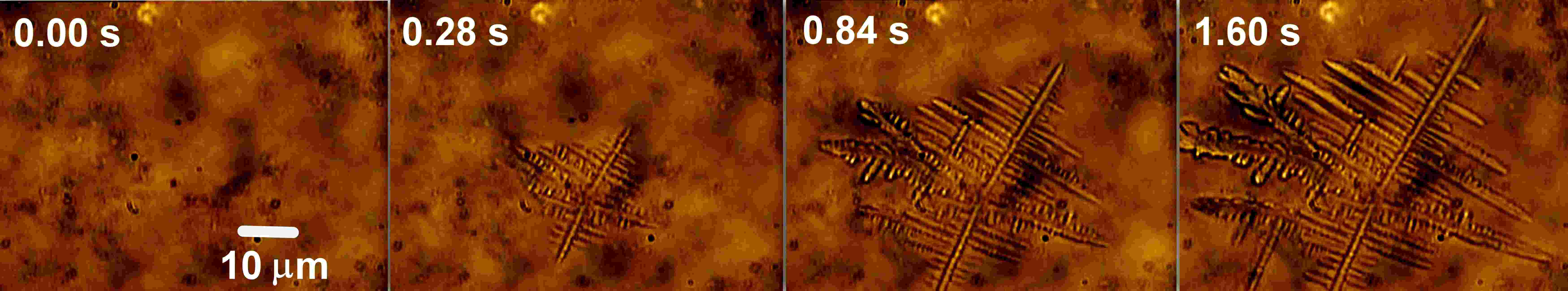}
\caption{}
\end{subfigure}
\caption{Time evolution of growing dentritic patterns in (a) graphite in PBS+Agarose, (b) SWNT in PBS+BSA, (c) MWNT in PBS+Gelatine. (In each of these case irradiation was by low-power (2 mW) laser light of wavelength 1064 nm), (d) SWNT in PBS+Gelatine 
and (e) MWNT in PBS+Agarose (in the last two cases 5 mW laser of the same wavelength has been used).}
\label{fig:images}
\end{figure}

We first consider the results obtained for three different solutions
(agarose, BSA and gelatine in PBS) and with three different absorbers (graphite,
SWNT and MWNT) for each solution. The patterns were recorded
using two different laser powers, 2 mW and 5 mW, and also under natural drying
(no laser) conditions. Figure~\ref{fig:images} shows
time evolution of patterns for selected solution and absorber combinations at 2 mW laser power.
As is obvious, very complex dendritic patterns are obtained and, remarkably, the growth of these patterns occurs on very fast (ms) time scales. The complexity of the patterns is most likely due to the anisotropy of the
absorber itself and also of the biological molecules present in the liquid.

In previous work~\cite{Bas} we had studied various aspects of
this type of dendritic growth, including measurement of the speed at which the tip grows as
a function of time and of the radius of curvature of this tip at every time step. We had shown that the product of the velocity and the square of the radius quickly decreases to a
value which remains constant at longer time scales, a hallmark of microscopic
solvability theory.

In this work we focus only on the measurements of the tip velocity
as a function of time and carry out this analysis for various systems that we have 
studied with the aim of testing the validity of our initial findings for diverse systems.

\begin{figure}
\centering
\begin{subfigure}{0.3\textwidth}
\includegraphics[width=\textwidth]{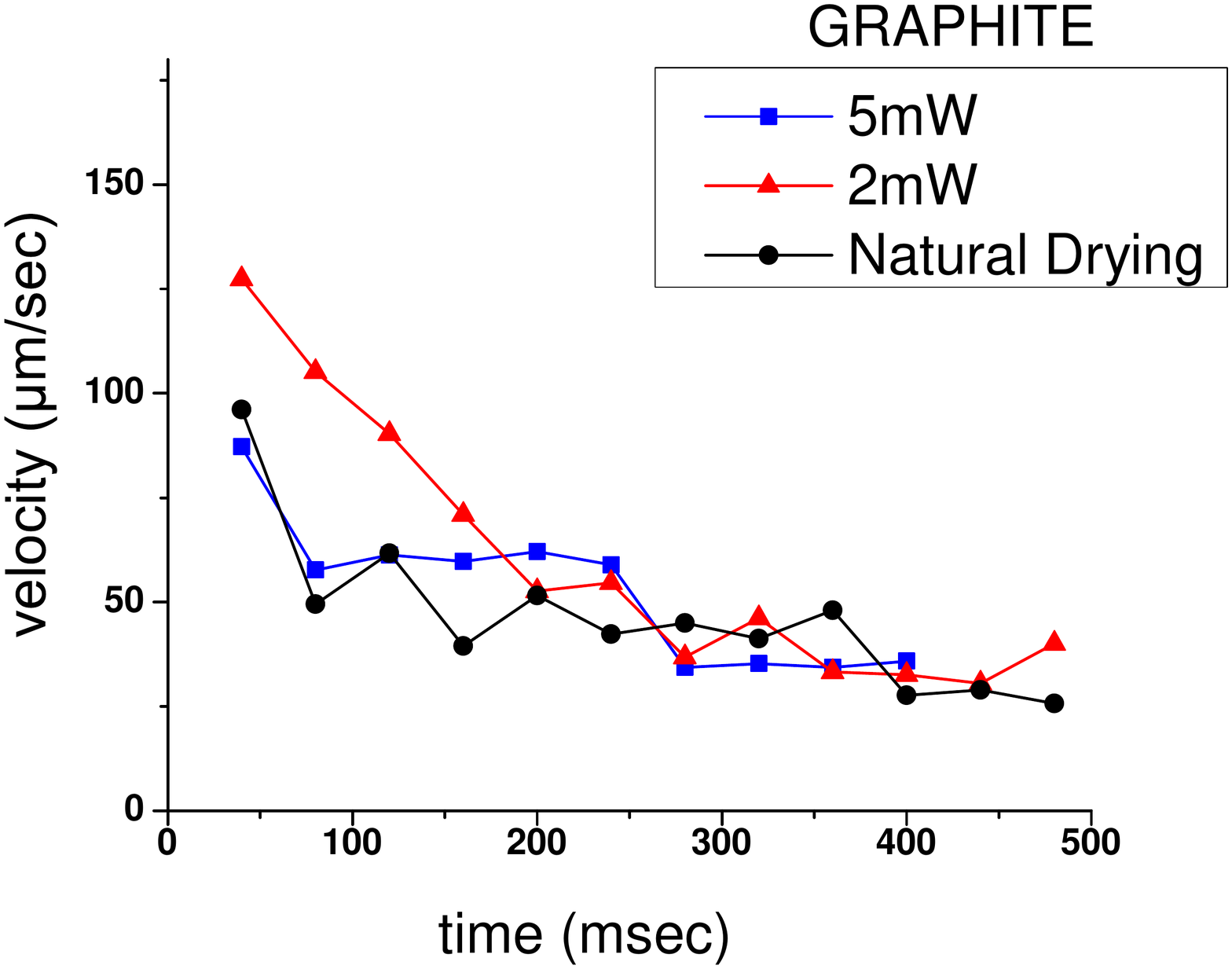}
\caption{}
\end{subfigure}
\begin{subfigure}{0.3\textwidth}
\includegraphics[width=\textwidth]{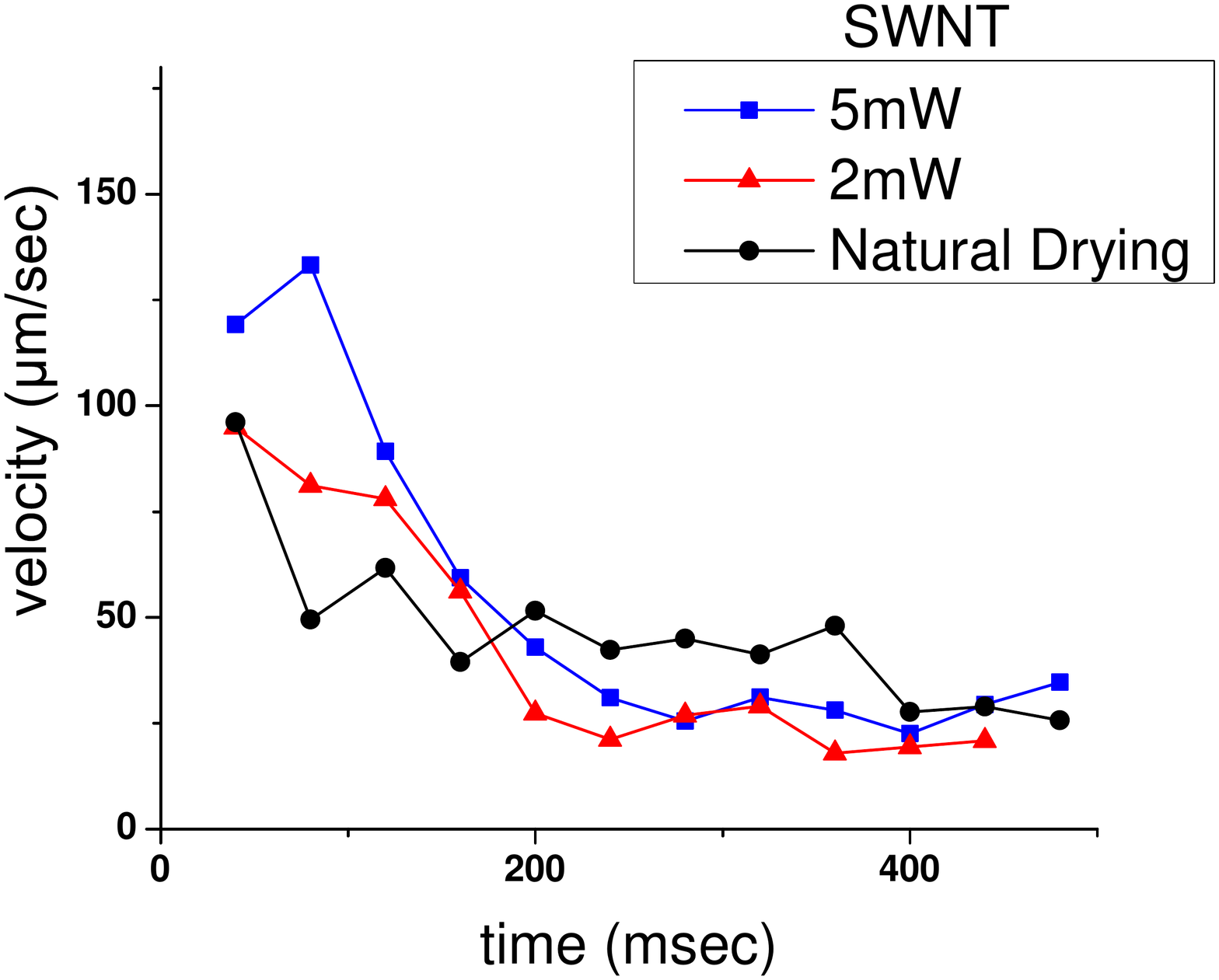}
\caption{}
\end{subfigure}
\begin{subfigure}{0.3\textwidth}
\includegraphics[width=\textwidth]{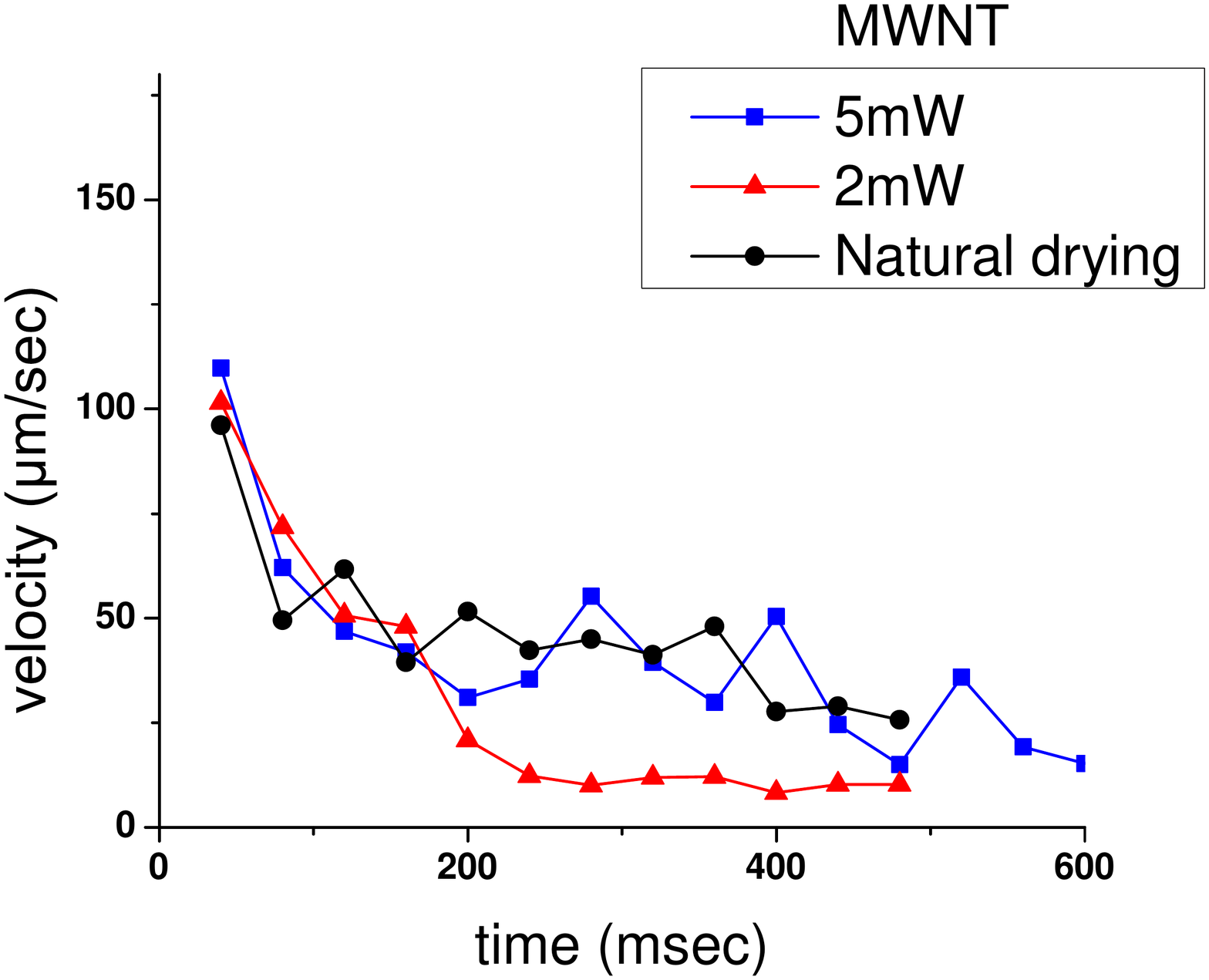}
\caption{}
\end{subfigure}
\caption{Dependence of tip velocity on time for PBS-Agarose solution and different
absorbers: (a) graphite, (b) SWNT, and (c) MWNT.}
\label{fig:agar}
\end{figure}

\begin{figure}
\centering
\begin{subfigure}{0.3\textwidth}
\includegraphics[width=\textwidth]{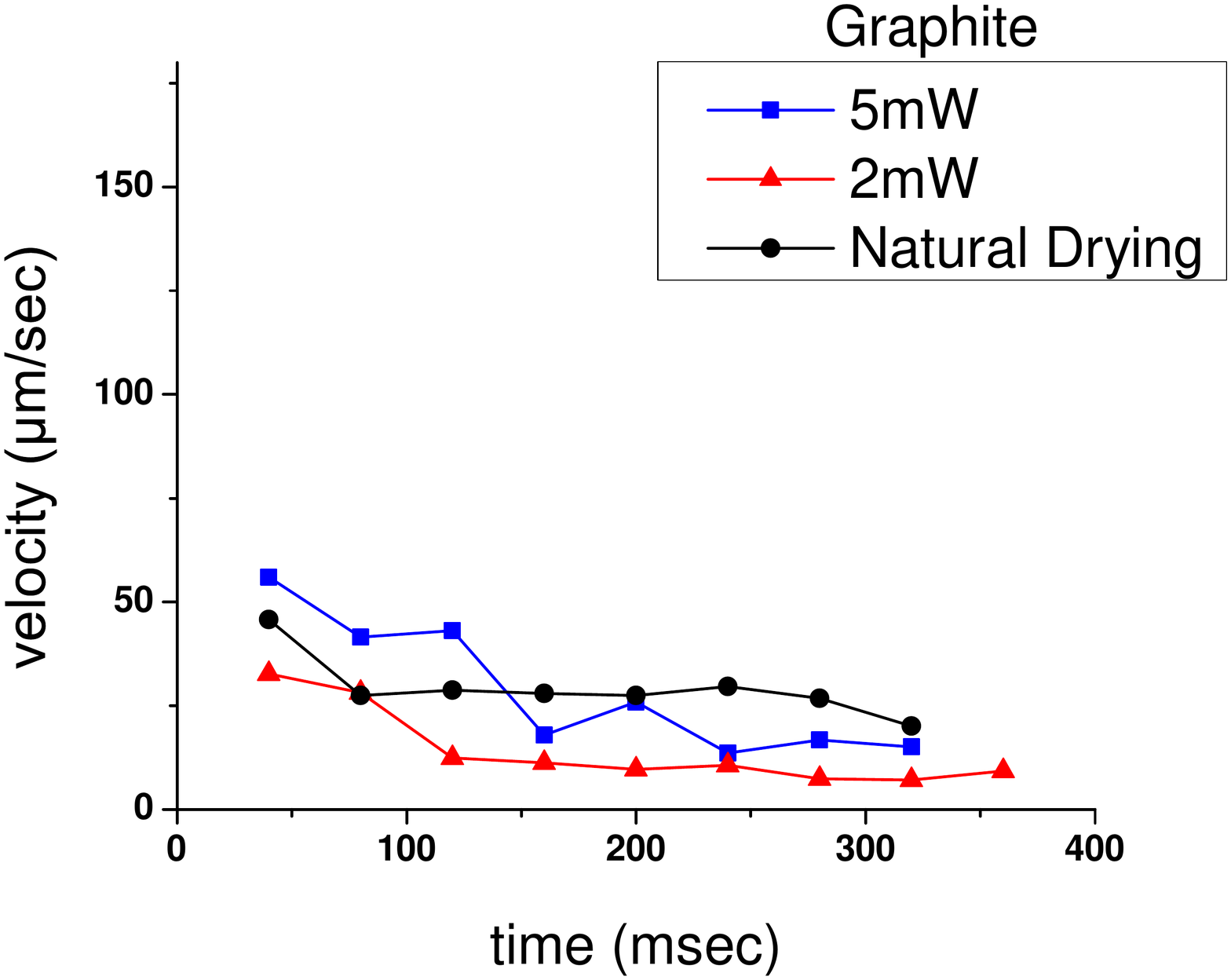}
\caption{}
\end{subfigure}
\begin{subfigure}{0.3\textwidth}
\includegraphics[width=\textwidth]{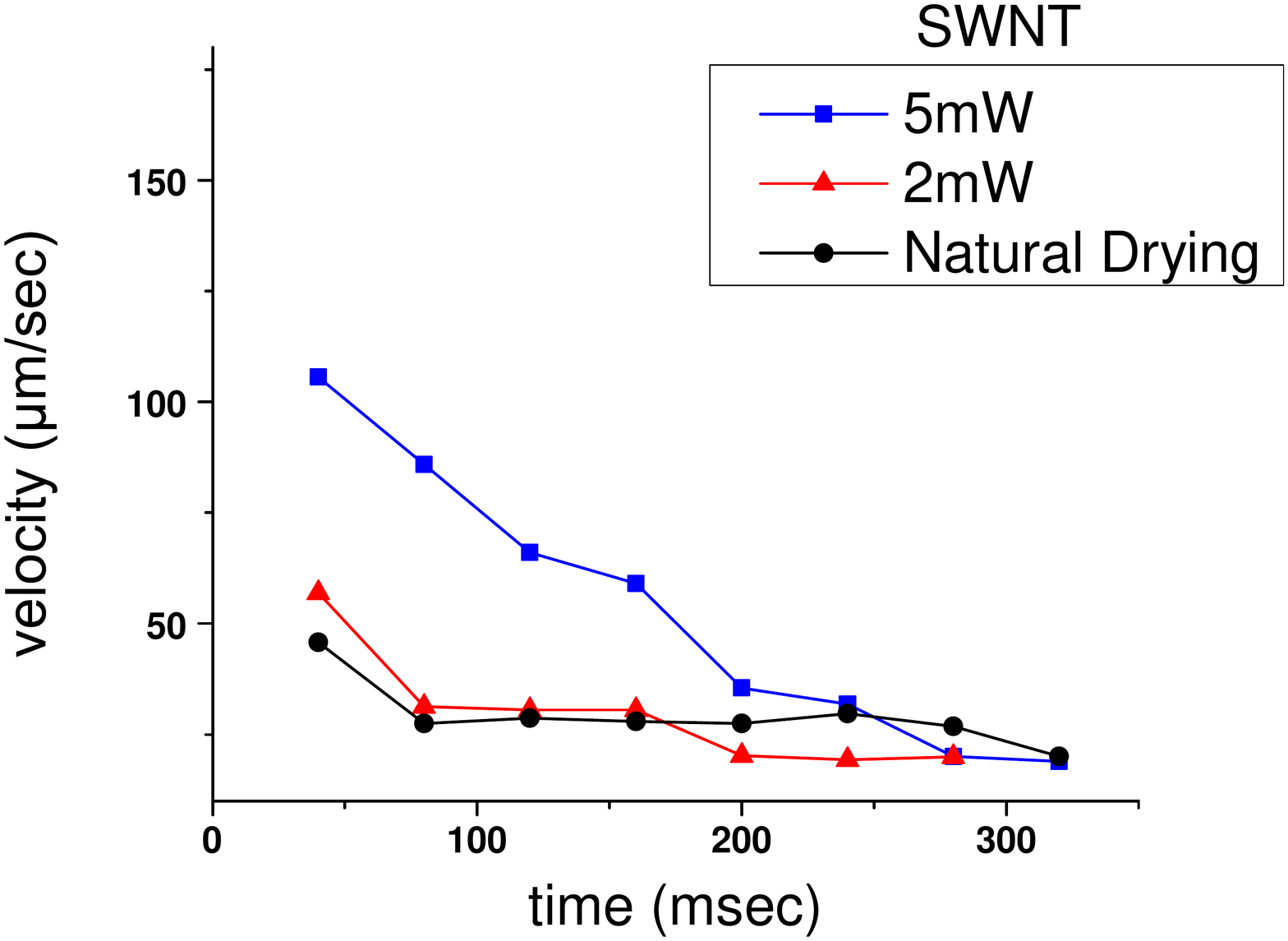}
\caption{}
\end{subfigure}
\begin{subfigure}{0.3\textwidth}
\includegraphics[width=\textwidth]{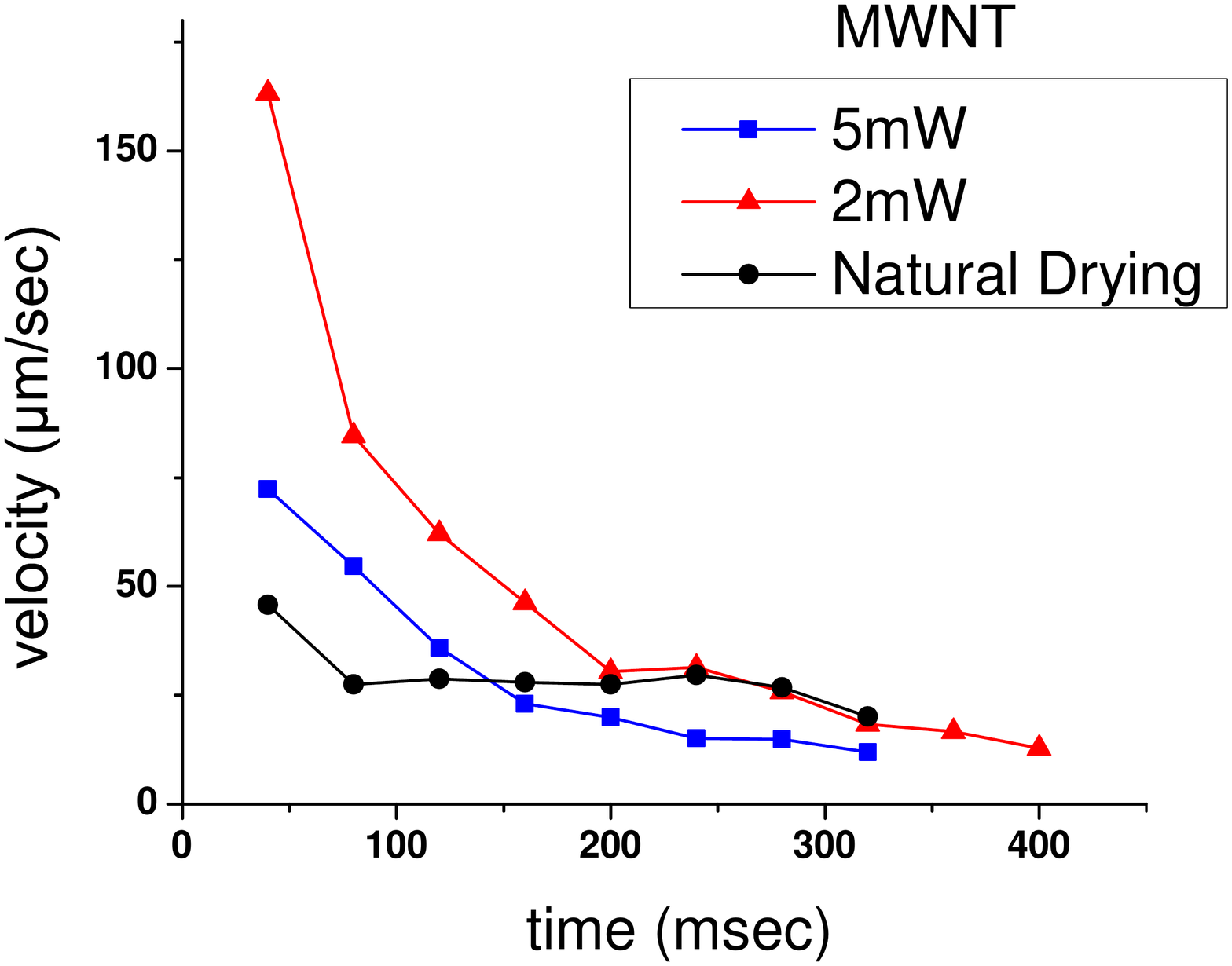}
\caption{}
\end{subfigure}
\caption{Dependence of tip velocity on time for PBS-Gelatine solution for different
absorbers: (a) graphite, (b) SWNT, and (c) MWNT.}
\label{fig:gelat}
\end{figure}

Figures~\ref{fig:agar} and \ref{fig:gelat} depict the functional dependence on time of the velocity of
a chosen tip for PBS in agarose and gelatine for
0 mW, 2 mW and 5 mW laser powers (the 0 mW data pertains to natural drying conditions). We see accelerated growth in the 
beginning phase of the dynamics followed by slow decrease in tip velocity as time progresses.
The observed temporal dependence of tip velocity can
be readily rationalized from the Stefan's boundary condition in equation~\ref{eq:bc}
which decides the velocity of the interface. When the interface is near
the absorber, even though the amount of heat is large, the temperature gradient
in the solid phase is very small. This leads to the large velocity of the
interface. As the interface moves away, the temperature gradient in the solid
phase increases and, consequently, the overall velocity comes down.

Another observation that is counterintuitive is that the
velocity of the tip with non-zero laser power crosses the value of the velocity obtained in the case of natural drying. When the laser is on, extra energy is pumped into the fluid which drives the growth, leading to large initial  speeds. Hence it is natural to expect that the velocity of the tip in this case will always be larger than that in the case when there is no laser radiation. It would, therefore, be expected that the velocities will approach each other only
asymptotically for large times. But what is observed in our experiments is that the velocity
of the tips driven by laser power actually reduce to values that are less than  those obtained without the laser light 
at finite times. This observation which was first reported in~\cite{Bas}
for BSA in PBS for SWNT and is now confirmed in the present experiments on different solutions and
with different absorbers. All the data presented in Figs.~\ref{fig:agar} 
and \ref{fig:gelat} demonstrate this fact.

It is possible to understand this observation mathematically from Eq.~(\ref{eq:bc}) which
suggests that the velocity of the tip is proportional to the difference
in the temperature gradients in the liquid and in the solid phase. When
there is no heat source present the temperature inside the solid phase
is almost the same everywhere and, hence, there is negligible temperature gradient as far as the solid is concerned. As a result, the velocity in this case is decided essentially by the temperature gradient in the fluid. But in the presence of laser light, energy
is transferred away from the source through the solid phase; this necessarily creates a temperature gradient there. Also, the sign of the
temperature gradients in the solid and the liquid need to be the same as
heat is being diffused away from the centre. Since Eq.~(\ref{eq:bc}) makes clear that 
the tip velocity is the difference of two gradients, it can be expected to assume values that lie below
the value for natural drying when the gradient in the fluid approaches
a steady state value.

It is also possible to understand this observation in a physical sense. When the optical 
energy from the laser is being supplied there has to be, as already noted, a temperature gradient in
the solid phase. This is needed for the diffusion of heat to occur. This
implies that when the interface moves further away from the source, the temperature of the 
newly created part of the solid has to be raised above the melting
temperature, implying that some energy is stored in the solid phase
as the interface moves away. Of course, when the interface is near
the source, the absorber itself supplies enough energy for the temperature to be raised to the necessary level and to support the accelerated growth that is observed. However, it is possible to imagine the situation when the interface is away from the absorber and the energy from the absorber
is insufficient to maintain the necessary temperature gradient. Under such circumstances, a part of the energy released from the latent heat may be
utilized for this purpose and stored in the newly created part of the 
solid as the interface progresses. As a result, less heat
is transferred to the fluid, and this has the effect of slowing down the interface. Clearly,
in the case of natural drying, since no energy needs to be stored, all of the latent heat is dissipated into the fluid and that is why the interface in the case of natural drying has a larger velocity in this regime.

Of course, as time progresses and the tip travels farther from the source, it would be expected that the velocities for different values of incident laser power would asymptotically approach each other and approach the value obtained in the case of natural drying. But, as is obvious from our discussion, such an approach to the natural drying value will be from {\em below} rather than above as one would have expected \emph{a priori}.

\section{Summary}\label{sec:dis}

We have presented here results from an exhaustive set of experiments that we conducted in which low powered laser light is made to irradiate an efficient absorber which transfers energy to the fluid in which
it is immersed. This gives rise to complex drying branching patterns. These patterns were studied in~\cite{Bas} for the first time and remarkably accelerated growth was demonstrated. Such accelerated growth of dendritic patterns is confirmed in the present studies conducted on different biological samples in different liquid environments using three different types of absorbers. We have measured the temporal behaviour of the tip velocity as well as of $vr^2$, where $v$ is the tip velocity and $r$ is the radius of curvature of
the tip of the observed branched patterns. We have found that in all cases the
tip velocity of the laser driven pattern decreases below the value obtained in the case of natural drying.
This counterintutive observation has been confirmed by carrying 
out experiments in solution of agarose, BSA and gelatine in PBS, using three different types of absorbers: graphite, SWNTs, and MWNTs. The crossover of tip velocities in all
the systems studied. We have rationalized our observations using mathematical and physical arguments but it is clear that our results open prospects for further theoretical work that will yield more quantitative insights into the complex dynamics. On the experimental front our work points to the necessity of making further observations long the lines reported in ~\cite{Bas} regarding the validity of the microscopic
solvability theory. It was observed there that the product $vr^2$,
which is expected to be constant from the microscopic solvability
theory, deviates from a constant value on very short time scales.
It would be interesting to confirm this breakdown of the theory
using other predictions of the theory, such as the coefficient 
of the fourth power in the dendrite shape.

\acknowledgments

We thank the Department of Science and Technology for support to JAD under the Women Scientists Scheme and for the J. C. Bose National Fellowship to DM.


%
%

\end{document}